\documentclass{article}%
\usepackage{amssymb}
\usepackage{amsmath}%
\usepackage{amsfonts}%
\usepackage{graphicx}
\begin{document}

\begin{center}
{\large Are elementary particles point-like objects?\medskip}
\end{center}

P. Brovetto$^{\dag}$, V. Maxia$^{\dag}$ and M. Salis$^{\dag\ddag}$
\footnote{Corresponding Author e-mail: masalis@unica.it}

$^{\dag}$On leave from Istituto Nazionale di Fisica della Materia - Cagliari, Italy

$^{\ddag}$Dipartimento di Fisica - Universit\`{a} di Cagliari, Italy\medskip

\textbf{Abstract - }Simple arguments based on uncertainty principle and
Dirac's equation are examined which show that electron behaves either as a
point-like charge or as an extended distribution according as high- or
low-energy experiments are considered.

Keywords: elementary particles, Dirac equation, Zitterbewegung theory.

PACS: 12.90.+b; 03.65.P.\medskip

Nowadays many physicists agree that elementary particles are point-like
objects. This opinion is based mostly on results of late experiments on
high-energy electron-positron collisions. These experiments show that the
electron size is less than $10^{-16}$ cm, that is, more than three orders of
magnitude smaller than the classical electron radius. In these experiments,
the collision time $\tau,$ as determined in the centre of mass of two
opposite-charge electrons moving with velocities $+v$ and $-v$, can be roughly
identified with the ratio between impact parameter $b$ and electron
velocities, that is, $\tau\simeq b/v$ $\left[  1\right]  .$ The impact
parameter in direction orthogonal to velocities is assumed equal to the
electron size, that is, $b_{\bot}=$ $10^{-16}$ cm. But $b_{\shortparallel
}=b_{\bot}\left(  1-\beta^{2}\right)  ^{1/2}$, owing to Lorentz contraction in
direction parallel to velocities. With electrons of energy $45$ GeV, we get
$\left(  1-\beta^{2}\right)  ^{1/2}=m_{e}c^{2}/\left(  45\text{ GeV}\right)
=1.1\cdot10^{-5}$ which, by letting $v=c$, leads to $\tau\simeq
b_{\shortparallel}/c=10^{-16}\left(  1-\beta^{2}\right)  ^{1/2}/c=3.7\cdot
10^{-32}$ s. This exceedingly small value of $\tau$ shows that the
electron-positron collision dealt with is indeed a very fast process. The
situation, of course, is quite different in low-energy experiments, as for
instance in atomic spectroscopy which involves long-living electron states. In
general, a high energy entails a "fast" process while a low energy entails a
"slow" process.

In order to decide whether electron shows a point-like nature also in
low-energy experiments, it is convenient to write the energy-time uncertainty
principle in the dimensionless form%
\begin{equation}
\frac{\delta w}{m_{e}c^{2}}\cdot\frac{\delta t}{T_{Z}}\simeq1,\label{aaa}%
\end{equation}
$m_{e}$ standing for the electron rest mass, and%
\begin{equation}
T_{Z}=\frac{h}{m_{e}c^{2}}=8.1\cdot10^{-21}\text{s}\label{bbb}%
\end{equation}
for twice the so-called Zitterbewegung period. As for the meaning of this
quantity, let us briefly recall some issues of Dirac's equation concerning
electron velocity $\left[  2,3,4,5\right]  $. According to this equation,
expected electron velocity on $x$-axis is $\left[  3\right]  $
\begin{equation}
\left\langle \overset{\cdot}{x}\left(  t\right)  \right\rangle =%
%TCIMACRO{\diiint }%
%BeginExpansion
{\displaystyle\iiint}
%EndExpansion
\sum_{k=1}^{k=4}\psi_{k}^{\ast}\left(  \overrightarrow{r},t\right)  \left(
-c\mathbf{\alpha}_{1}\right)  \psi_{k}^{{}}\left(  \overrightarrow
{r},t\right)  d^{3}\ \overrightarrow{r},\label{ccc}%
\end{equation}
where $\psi_{k}$ are the spinor components. Fourier's expansion allows us to
write%
\[
\psi_{k}^{{}}\left(  \overrightarrow{r},t\right)  =%
%TCIMACRO{\diiint }%
%BeginExpansion
{\displaystyle\iiint}
%EndExpansion
\left\{  a_{k}\left(  \overrightarrow{p}\right)  \exp\left[  \left(  2\pi
i/h\right)  \left(  Wt-\overrightarrow{p}\cdot\overrightarrow{r}\right)
\right]  +\right.
\]

\begin{equation}
\left.  +b_{k}\left(  \overrightarrow{p}\right)  \exp\left[  \left(  2\pi
i/h\right)  \left(  -Wt-\overrightarrow{p}\cdot\overrightarrow{r}\right)
\right]  \right\}  d^{3}\ \overrightarrow{p}\label{ddd}%
\end{equation}
where $W$ means the electron energy and $a_{k}$, $b_{k}$ the amplitudes of
positive and negative energy states, respectively. The latter are essential
because it was shown that without their contribution electrons cannot
originate Thomson scattering of light. This scattering appears as a sort of
resonance of the quantum jump of energy $2m_{e}c^{2}$ between positive and
negative energy states $\left[  6\right]  $. Substituting eq. (\ref{ddd}) into
eq.(\ref{ccc}), and integrating over $\overrightarrow{r}$ originates terms
$\mathbf{-}ca_{k}^{\ast}\mathbf{\alpha}_{1}a_{k}$ and $-cb_{k}^{\ast
}\mathbf{\alpha}_{1}b_{k}$ which lead to the Ehrenfest theorem as in
non-relativistic quantum mechanics. Terms $\mathbf{-}ca_{k}^{\ast
}\mathbf{\alpha}_{1}b_{k}\exp\left(  -4\pi iWt/h\right)  $ $+$ H.C. are also
found which couple positive with negative energy states. They represent
oscillations of velocity of angular frequency $4\pi W/h$. Since we are dealing
with low energy electrons, we assume that $W$ barely exceeds $m_{e}c^{2}$. So,
by putting $W=m_{e}c^{2}$, we have from eq. (\ref{bbb}) $W/h=1/T_{Z}$.
Consequently, with obvious meaning of labels $E$ and $Z$, the expected
velocity can be written as%

\begin{equation}
\left\langle \overset{\cdot}{x}\left(  t\right)  \right\rangle =\left\langle
\overset{\cdot}{x}\left(  t\right)  \right\rangle _{E}+\left\langle
\overset{\cdot}{x}\left(  t\right)  \right\rangle _{Z}.\label{eee}%
\end{equation}
Component $\left\langle \overset{\cdot}{x}\left(  t\right)  \right\rangle _{E}
$ is not at issue. As for component $\left\langle \overset{\cdot}{x}\left(
t\right)  \right\rangle _{Z}$, by performing some transformatios, we get%

\begin{equation}
\left\langle \overset{\cdot}{x}\left(  t\right)  \right\rangle _{Z}=c%
%TCIMACRO{\diiint }%
%BeginExpansion
{\displaystyle\iiint}
%EndExpansion
A_{1}\cos\left(  4\pi\frac{t}{T_{Z}}+\varphi_{1}\right)  d^{3}%
\ \overrightarrow{p},\label{fff}%
\end{equation}
where amplitude $A_{1}$ and phase $\varphi_{1}$ are real quantites, depending
on $\overrightarrow{p},$ which can be evaluated in function of Fourier's
amplitudes $a_{k}$ and $b_{k}$. Integration over time, yields in turn%

\begin{equation}
\left\langle x\left(  t\right)  \right\rangle _{Z}=\frac{\lambda_{C}}{4\pi}%
%TCIMACRO{\diiint }%
%BeginExpansion
{\displaystyle\iiint}
%EndExpansion
A_{1}\sin\left(  4\pi\frac{t}{T_{Z}}+\varphi_{1}\right)  d^{3}%
\ \overrightarrow{p},\label{ggg}%
\end{equation}
$\lambda_{C}=cT_{Z}=h/m_{e}c$ standing for Compton's wavelength. Taking also
into account components $\left\langle y\left(  t\right)  \right\rangle _{Z}$
and $\left\langle z\left(  t\right)  \right\rangle _{Z}$, it follows that each
element $dp_{x}dp_{y}dp_{z}$ of momentum space is associated with a closed
trajectory formed by electron oscillations on $x,y,z$ axes. Adding up these
trajectories, a complex dynamic substructure is originated in space around the
electron centre of mass $\left[  7\right]  $. So far an exhaustive treatment
on this matter has not been implemented, due perhaps to the difficulty of
rightly define the low-energy spinor $\psi_{k}$ in eq. (\ref{ccc}).

Coming back to eq. (\ref{aaa}), we point out that in low-energy experiments we
have in general\ $\delta w/m_{e}c^{2}\ll1$. Indeed, in atomic spectroscopy
energies are determined with accuracies so good that the Rydberg constant,
$R_{H}=13.5056981$ eV, is known with a precision of seven decimal digits. This
entails $\delta w<10^{-7}$ eV and $\delta w/m_{e}c^{2}<2\cdot10^{-13}$.
Accordingly, we obtain from eq. (\ref{aaa}) $\delta t/T_{Z}>5\cdot10^{12}$. It
follows that oscillations considered in eqs. (\ref{ggg}) and (\ref{fff}) are
not observable since indetermination in time largely exceeds oscillation
period. Consequently, time must be eliminated in previous equations so that
oscillations are changed into a probability density distributed on the
electron trajectories \footnote{This is like what occurs when dealing with a
classic oscillator $x=Asin\left(  2\pi t/T\right)  $. Probability\ that the
oscillating particle is found in space between $x$ and $x+dx$ is $dP=2dt/T$,
$\ dt$ standing for the time required to cross $dx$. We have thus
$dP/dx=2/\left(  \overset{\cdot}{x}T\right)  $ which allows us to write
$dP/dx=1/\left(  \pi\sqrt{A^{2}-x^{2}}\right)  $ (see Ref. $\left[  8\right]
$). A \ simple way to grasp the argument at issue is as follows. In
Bohr-Sommerfeld mechanics atomic electrons move on closed trajectories. In
quantum mechanics these trajectories are replaced by time-independent clouds
of probability just as a consequence of energy-time uncertainty principle.}.
In our opinion, this argument, although only qualitative, is sufficient to
conclude that in low-energy experiments electron behaves as an extended charge
of definite size.

This outcome is not surprising because in quantum physics electron cannot be
regarded as an "absolute" entity. Electron, on the contrary, is merely an
"observable" object. It follows that in actual experiments different sides of
electron can be probed, depending on the experiment considered. So, electron
is point-like in fast collisions at high energy ($\tau\ll T_{Z}$) but is
extended in low-energy atomic spectroscopy. Since all elementary particles are
fermions represented by Dirac's equation, this results should be applied to
all particles barring neutrinos. Indeed, their yet unknown vanishingly small
mass prevents eq. (\ref{bbb}) to be applied. We point out, finally, that the
rest-mass of the extended electron might be ascribed to its own
electromagnetic field without run into divergent quantities as with the
point-like option. Some conjectures about this matter are reported in Ref.
$\left[  8\right]  \medskip$

\textbf{References}

$\left[  1\right]  $ E. Fermi, \textit{Nuclear Physics} (University of Chicago
Press, 1955) Ch. II.

$\left[  2\right]  $ E. Schr\"{o}dinger, Sitzungsb. Preuss. Akad. Wiss. Phys.
Math. Kl. \textbf{24}, 418 (1930); \textbf{3}, 1 (1931).

$\left[  3\right]  $ L. De Broglie, \textit{L'Electron Magnetique} (Hermann,
Paris, 1934) Ch. XXI.

$\left[  4\right]  $ P.A.M. Dirac, \textit{The Principles of Quantum
Mechanics} (Oxford, 1958) \S \ 69.

$\left[  5\right]  $ B.R. Holstein, \textit{Topics in Advanced Quantum
Mechanics} (Addison-Wesley, 1992).

$\left[  6\right]  $ E. Fermi, Rev. Mod. Phys. \textbf{4}, 120 (1932).

$\left[  7\right]  $ A.O. Barut and A.J. Bracken, Phys. Rev. D \textbf{23,}
2454 (1981).

$\left[  8\right]  $ P. Brovetto, V. Maxia and M. Salis -
arXiv:quant-ph/0512047 (10 Apr 2006).
\end{document}